\documentclass[conference]{IEEEtran}

\usepackage{floatflt}

\usepackage{color}
\usepackage{cite}

\usepackage{graphicx}
\usepackage{amsmath}
\usepackage{amsthm}
\usepackage{amssymb}

\usepackage{verbatim}

\usepackage{psfrag,array}

\usepackage{subfigure}

\usepackage{stfloats}

\usepackage{todonotes}

\usepackage{amsmath}
\usepackage{epstopdf}
\usepackage{algorithmic}

\newcommand{\p}[1]{\mathop{\mbox{\it p} } }

\renewcommand{\vec}[1]{\ensuremath{\boldsymbol{#1}}}
\newcommand{\be}{\begin{equation}}
\newcommand{\ee}{\end{equation}}
\newcommand{\ba}{\begin{array}}
\newcommand{\ea}{\end{array}}
\newcommand{\bea}{\begin{eqnarray}}
\newcommand{\eea}{\end{eqnarray}}
\newcommand{\bean}{\begin{eqnarray*}}
\newcommand{\eean}{\end{eqnarray*}}

\definecolor{white}{rgb}{1,1,1}

\newtheorem{property}{Property}

\newtheorem{corollary}{Corollary}
\newtheorem{remark}{Remark}

\begin{document}

\title{The Potential of Using Large Antenna Arrays on Intelligent Surfaces}

\author
{
\begin{tabular}{c}
 Sha Hu, Fredrik Rusek, and Ove Edfors  $\quad\quad$ \\
 Department of Electrical and Information Technology $\quad\quad$\\
 Lund University, Lund, Sweden $\quad\quad$\\
 \{firstname.lastname\}@eit.lth.se $\quad\quad$
\end{tabular}
}

\maketitle

\begin{abstract}
In this paper, we consider capacities of single-antenna terminals communicating to large antenna arrays that are deployed on surfaces. That is, the entire surface is used as an intelligent receiving antenna array. Under the condition that the surface area is sufficiently large, the received signal after matched-filtering (MF) can be well approximated by an intersymbol interference (ISI) channel where channel taps are closely related to a sinc function. Based on such an approximation, we have derived the capacities for both one-dimensional (terminals on a line) and high dimensional (terminals on a plane or in a cube) terminal-deployments. In particular, we analyze the normalized capacity $\bar{\mathcal{C}}$, measured in nats/s/Hz/m$^2$, under the constraint that the transmit power per m$^2$, $\bar{P}$, is fixed. We show that when the user-density increases, the limit of $\bar{\mathcal{C}}$, achieved as the wavelength $\lambda$ approaches 0, is $\bar{P}/(2N_0)$ nats/s/Hz/m$^2$, where $N_0$ is the spatial power spectral density (PSD) of noise. In addition, we also show that the number of signal dimensions is $2/\lambda$ per meter deployed surface for the one-dimensional case, and $\pi/\lambda^2$ per m$^2$ deployed surface for two and three dimensional terminal-deployments.
\end{abstract}

\section{Introduction}
We envision a future where man-made surfaces become electronically active, enabling wireless communication, wireless charging, and remote sensing, making the physical environment \lq\lq{}intelligent\rq\rq{} and interactive. This makes it possible to fulfill the most grand visions for the Internet of Things \cite{IOT}, where many billions of devices are expected to be connected to the Internet.

\begin{figure}[t!]
\begin{center}
\vspace*{-0mm}
\hspace*{-2mm}
\scalebox{0.63}{\includegraphics{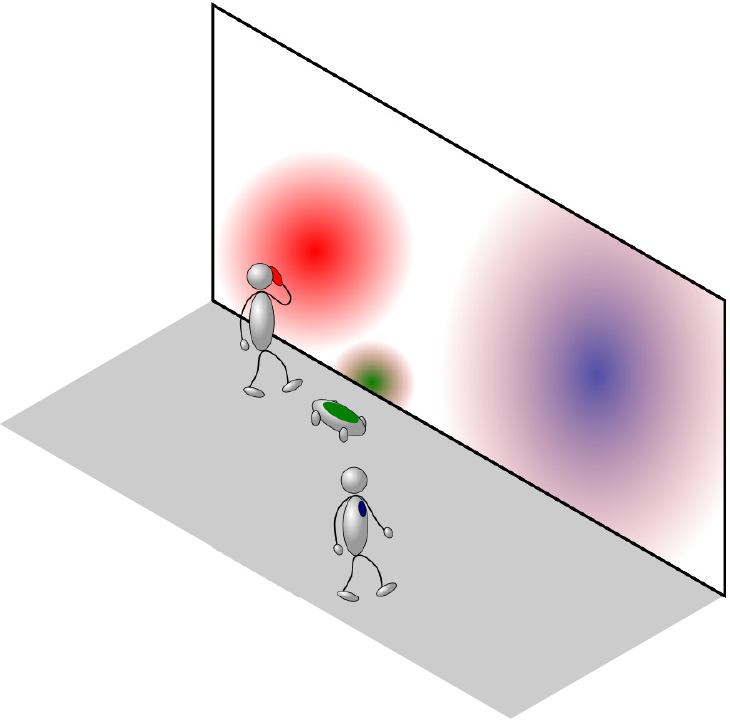}}
\vspace*{-2mm}
\caption{\label{fig0}Three users communicating with a large intelligent surface.}
\vspace*{-6mm}
\end{center}
\end{figure}

With intelligent surfaces we mean surfaces in the physical environment that are electromagnetically active, where each part of the surface can transmit and receive electromagnetic fields. Being able to carefully control these fields makes it possible to tightly focus energy in three-dimension space both for transmission and reception so the surfaces will bring entirely new capabilities both for communication, sensing and control of the electromagnetic environment; see Fig. \ref{fig0} for an illustration of the intelligent surface concept. Intelligent surfaces can be seen as the natural evolution of the massive MIMO concept \cite{MM}, but taken to the extreme. The benefits of Massive MIMO are today well
understood but there has only been limited previous attempts to take the disruptive step from a large number of antennas on a base station (as in massive MIMO) to using surfaces in an entire physical environment as \lq\lq{}the antenna\rq\rq{}.

One attempt in this direction is done by the Berkeley ewallpaper project, where the ultimate vision is to fabricate wall papers that are electromagnetically active and has built-in processing power \cite{ewall}. However, no analysis has been carried out on information transfer capabilites of intelligent surfaces. Rather, the efforts have been directed towards implementation and hardware aspects of intelligent surfaces. 

In this paper we take a first look at the information transfer capabilities of an intelligent surface. In particular, we  show that for every m$^2$ deployed surface area, $\pi/\lambda^2$ users can be spatially multiplexed. We also demonstrate, through numerical simulation, that a fairly small intelligent surface can yield per-user capacity to around one hundred users in a medium sized room virtually as well as if only one user was present.

The rest of the paper is organized as follows. In Sec. II we describe the received signal model for intelligent surfaces and introduce a sinc approximation for analytical tractability. In Sec. III we analyze the capacities for both the optimal and MF receivers with one-dimension terminal-deployment. In Sec. IV we derive the number of independent signal dimensions both for two and three dimensional cases. Numerical results are presented in Sec. V, and Sec. VI summarizes the paper.

\section{System Model}
We consider the transmission from a number of autonomous single antenna terminals located in three-dimensional space to a large two-dimensional surface. Expressed in Cartesian coordinates, the surface is located at $-A\leq x \leq A$, $-B \leq y \leq B$, and $z=0$, while terminals are located at $z\!>\!0$ and at arbitrary $x$, $y$ coordinates. For analytical tractability we assume an ideal situation where no scatterers or reflections are present, yielding a perfect line-of-sight (LoS) propagation scenario. Each terminal is assumed to propagate an isotropic signal. The received signal at the surface at location ($x$,\,$y$), assuming a narrow-band system and ideal free-space propagation from the terminal to that point, corresponding to a terminal at location $(x,y,z) = (x_0,y_0,z_0)$ is spread across the surface according to
{\setlength\arraycolsep{0pt}
\bea s_{x_0,y_0,z_0}(x,y)&=&\frac{1}{2\sqrt{\pi}} \frac{\sqrt{z_0}}{\left(z_0^2+(x-x_0)^2+(y-y_0)^2\right)^{3/4}}\nonumber \\
&&\times \exp\!\left(\!-\frac{2\pi\jmath}{\lambda}\sqrt{z_0^2+(y-y_0)^2+(x-x_0)^2}\right)\!, \nonumber  \eea}
\hspace{-1.4mm}where $\lambda$ is the wavelength. 

Assuming $K$ terminals, where the $k$th terminal is located at $(x_k,y_k,z_k)$ and is transmitting the data symbol $a_k$, the total received signal at position $(x, y)$ on the surface is
\bea r(x,y) = \sum_{k=1}^K \sqrt{P_k} s_{x_k,y_k,z_k}(x,y)a_k +n(x,y),\eea
where $P_k$ is the transmit power of the $k$th terminal and $n(x,y)$ is zero-mean complex Gaussian noise, independent over $x$ and $y$, and with spatial power spectral density (PSD) $N_0$. 

Given the received signal across the surface, optimum processing includes applying a spatial correlator to each transmit signal, a procedure we will call matched filtering (MF),
\bea r_k = \int_{-A}^A\int_{-B}^B r(x,y)\sqrt{P_k}s_{x_k,y_k,z_k}^\ast(x,y)\mathrm{d}x \mathrm{d}y. \notag \eea
Assembling the notation into  a matrix formulation, we have that 
\bea \label{rk} \vec{r} =\vec{G}\vec{a} + \vec{w}, \eea
where the $(\ell, k)$th element $G_{\ell k}$ of matrix $\vec{G}$ equals $G_{\ell k}=\sqrt{P_\ell P_k}\phi_{\ell k},$ and
\bea \label{phi} \phi_{\ell k} =\int_{-A}^A\int_{-B}^Bs_{x_k,y_k,z_k}(x,y)s_{x_\ell,y_\ell,z_\ell}^\ast(x,y)\mathrm{d}x \mathrm{d}y. \eea
Moreover, with MF applied, the noise variables are zero mean but colored with covariance matrix $\mathbb{E}[\vec{w}\vec{w}^\mathrm{H}]=\vec{G}.$

In the rest of this paper, we assume equal terminal transmit powers $P_k\!=\!P$ and study the capability of the terminals to communicate with the surface. In particular, we put an emphasis on the number of independent dimensions per area unit of deployed surface that is possible to harvest. That is, we are interested in the ratio 
\bea \label{rho} \rho=\frac{\mathrm{rank}(\vec{G})}{V}, \eea
where $V$ is the volume specified by all the $K$ terminals that are deployed in a three-dimensional space. In one-dimensional deployment, all the terminals are along a line and $V$ is then the total length of the line, and in two-dimensional deployment, all the terminals are on a plane and $V$ is the area of the plane specified by all the $K$ terminals.

\subsection{Array Gain Considerations}
Let us consider the received power at the surface from an omni-directional antenna with power $P$ that is located at coordinates $x=0$, $y=0$ and $z=z_0$, that is, $z_0$ meters from the surface and perpendicular to its center. The total received power at the surface is then given by the diagonal elements $G_{kk}$ that equal
{\setlength\arraycolsep{2pt} \bea \label{Prx} G_{kk} &=& P\int_{-A}^A \int _{-B}^B |s_{0,0,z_0}(x,y)|^2 \mathrm{d}x\mathrm{d}y \nonumber \\
&=&\frac{P}{4\pi}\int_{-A}^A \int _{-B}^B \frac{z_0}{\left(z_0^2+x^2+y^2\right)^{3/2}} \mathrm{d}x\mathrm{d}y \nonumber \\
&=& P\nu,\eea}
\hspace{-1.4mm}with the variable $\nu$ ($0\!\leq\!\nu\!\leq\!1/2$) being
\bea \label{nu} \nu=\frac{1}{\pi}\tan ^{-1}\left(\frac{AB}{z_0\sqrt{A^2+B^2+z_0^2}} \right). \eea
Under the case that the surface is infinitely long, that is, $A\!=\!\infty$, the total received power at the surface is
\bea \label{Prx1} G_{kk}= \frac{P}{\pi}\tan ^{-1}\left(\frac{B}{z_0} \right).\notag \eea
Moreover, if the surface is also infinitely wide, that is, $B\!=\!A\!=\!\infty$, the total received power equals
\bea \label{Prx1} G_{kk} = P/2,\notag \eea
which makes intuitive sense, since half of the isotropically transmitted power from the terminal will reach the surface, while the other half propagates away from it.

This number should now be compared with the free-space path loss that would result from a single receive antenna at distance $z_0$, which equals $(\lambda/(4\pi z_0))^2 P.$ As this number is typically many orders of magnitudes smaller than $P/2$, we obtain, in addition to a possibly large value on the number of independent dimensions per volume unit (\ref{rho}), an impressive array gain.

\subsection{On the Approximation of an Integral for Large Surfaces} \label{Sec:approx}
\begin{figure}[t]
\begin{center}
\vspace*{-3mm}
\hspace*{-7mm}
\scalebox{.31}{\includegraphics{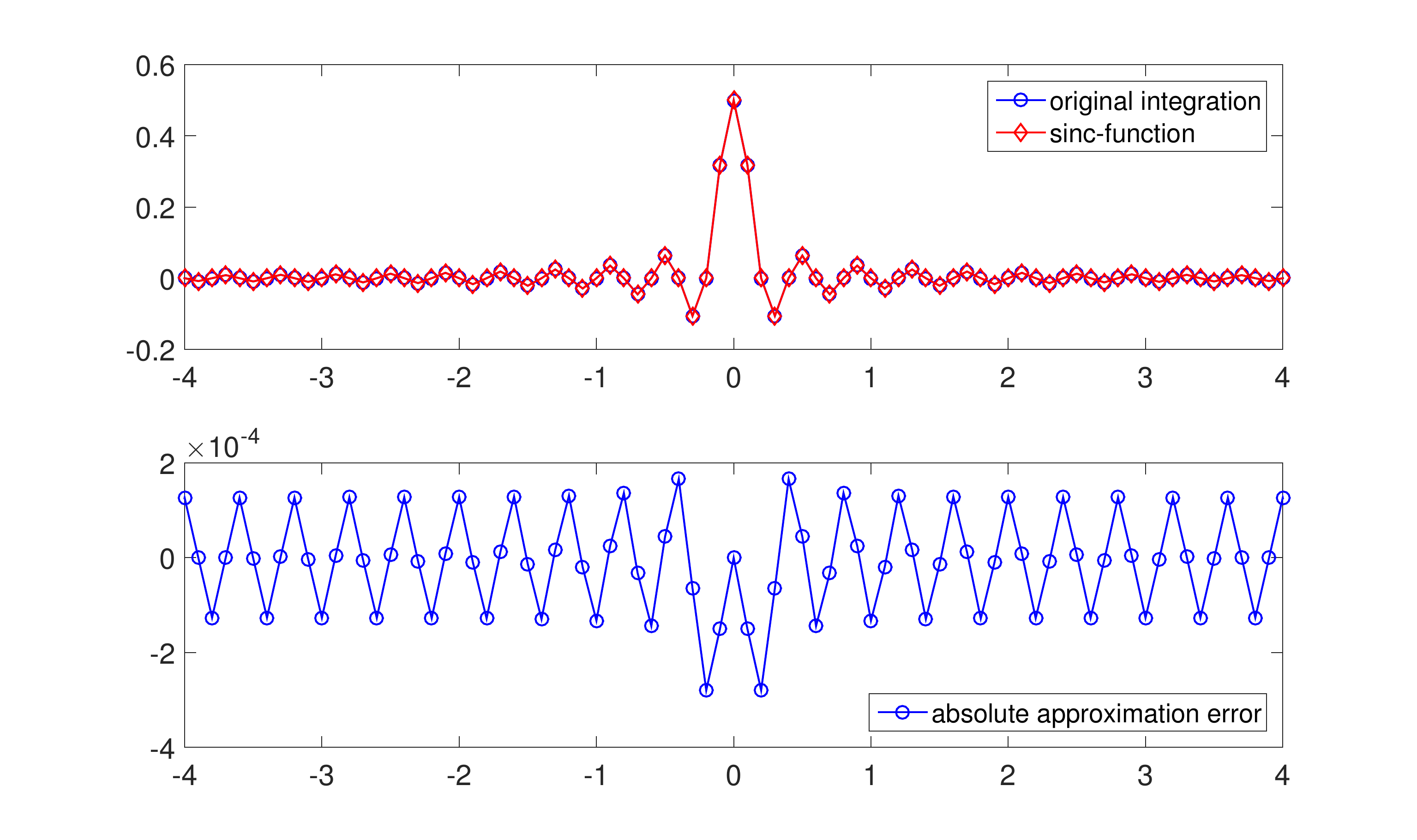}}
\vspace*{-9mm}
\caption{\label{fig1}The approximation of integration (\ref{eq1}) and sinc-function (\ref{eq2}) for $g(\Delta x)$ with $d\!=\!2$ and $\lambda\!=\!0.4$.}
\vspace*{-6mm}
\end{center}
\end{figure}

Working with large surfaces will result in the need of solving an integral to calculate $\phi_{\ell k}$ $(\ell\!\neq\!k)$ that, unfortunately, does not seem to have any closed form solution. However, there is a simple approximation to the integral that is remarkably tight. The integral of concern is the following
{\setlength\arraycolsep{0pt}
\bea \label{eq1} g&&(\Delta x) = \int_{-\infty}^{\infty}\left(z^2+x^2\right)^{-3/4}\left(z^2+(x+\Delta x)^2\right)^{-3/4} \nonumber \\
&& \times \exp\!\left(\!-\frac{2\pi\jmath}{\lambda}\left[\sqrt{z^2+x^2}\!-\!\sqrt{z^2+(x+\Delta x)^2}\right]\right)\mathrm{d}x,\quad \eea }
\hspace{-1.4mm}for some arbitrary $z\!>\!0$. This integral can be well approximated by 
\be \label{eq2} g (\Delta x) = \frac{2}{z^2}\mathrm{sinc}\left(\frac{2}{\lambda}\Delta x\right),\ee
where $\mathrm{sinc}(x)\!=\!\sin(\pi x)/(\pi x)$. In Fig. \ref{fig1} we depict $g(\Delta x)$ corresponding to (\ref{eq1}) and (\ref{eq2}) for $d\!=\!2$ and $\lambda\!=\!0.4$, respectively. As can be seen, the two curves are close to each other. With the approximation (\ref{eq2}), we can then analyze the capacity of the large surface in forthcoming sections.

\section{Capacity for One-Dimensional Case: Terminals on a Line}
\begin{figure}[b!]
\begin{center}
\vspace*{-6mm}
\hspace*{-1mm}
\scalebox{.34}{\includegraphics{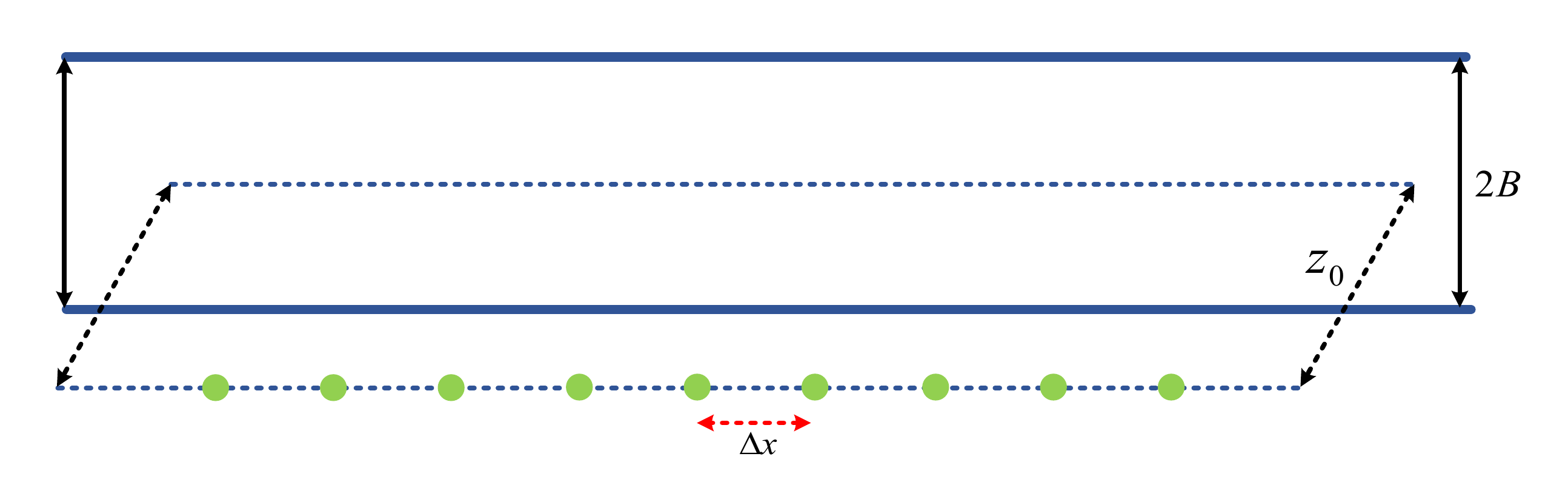}}
\vspace*{-9mm}
\caption{\label{fig2}Terminals on a line communicating to a infinitely long wall.}
\vspace*{-4mm}
\end{center}
\end{figure}

We start with one-dimension terminal-deployment and consider an infinitely long wall, i.e., $A\!=\!\infty$, where terminals are uniformly distributed on a line ($y_k\!=\!0$) with a distance $\Delta x$ between two adjacent terminals. Such a communication system is depicted in Fig. \ref{fig2}. Although an infinitely long wall and equi-distant terminal locations are unreasonable in practice, these assumptions are made for analytical tractability. General capacity results will be obtained, from which the general capacity behavior can be concluded. Numerical results on surfaces with finite sizes and random terminal positions will be given in Sec. V, which show that the numerical results are well predicted by the theoretical analysis.

For notational convenience, we define the ratio $\theta$ between the half wave-length and the terminal-distance as
 \bea \label{theta} \theta\!=\!\lambda/(2\Delta x). \eea
As we are assuming that an MF is applied as front-end, from (\ref{rk}) the received signal can be expressed as
\bea \label{mod1} r_k =\sum_{\ell=-\infty}^{\infty}g_{\ell}a_{k-\ell}+w_{k},\eea
where $a_k$ and $r_k$ are the transmitted and received signals for the $k$th user, and $w_k$ is additive noise with  correlation
\bea \label{corr} \mathbb{E}[n_{k}n_{k+l}^{\ast}]=g_\ell N_0. \notag \eea
The effective channel impulse response $g_{\ell}$ is real and can be approximated as 
\bea \label{gl} g_\ell =P\nu\mathrm{sinc}\left(\frac{2\ell\Delta x}{\lambda}\right),\eea
where $\nu$ is from (\ref{nu}).

\subsection{Capacity with Optimal Receiver}
After the MF and with model (\ref{mod1}), we can successively apply an optimal receiver, and the capacity [nats/s/Hz] of each terminal equals
 \bea \label{Copt} \mathcal{C}=\frac{1}{\theta}\int_{-\theta/2}^{\theta/2}\!\log\left(1+\frac{G(f)}{N_0}\right)\mathrm{d}f. \eea
Noticing that, $g_\ell$ are discrete samples of the sinc function with sampling rate $\theta$, and by the Poisson summation formula \cite{BZ97}, the frequency response $G(f)$ equals
\bea \label{Gf} G(f)=\theta P\nu\sum_{k=-\infty}^{\infty}G_0(f-k\theta), \eea
where $G_0(f)$ is the standard rectangular function (i.e., the Fourier transform of the sinc-function). To avoid that total transmit power per meter, denoted $\bar{P}$, grows without bounds when terminal density is increased, we constrain transmit power per terminal as
\bea \label{con1} P/\Delta x=\bar{P},\eea
and the capacity in (\ref{Copt}) can be explicitly computed. Defining two auxiliary variables
\bea \label{a} \alpha=1/\theta-\beta \;\;\mathrm{and}\;\; \label{b} \beta=\left\lfloor 1/\theta\right\rfloor, \eea
the capacity for the one-dimension case is given in Property~1.
\begin{property} 
The capacity (\ref{Copt}) equals
 {\setlength\arraycolsep{2pt} \bea \label{Copt3} \mathcal{C}&=& \alpha\log\left(1+\frac{(\beta+1) \lambda\bar{P} \nu}{2N_0}\right) \notag \\
 &&+(1-\alpha)\log\left(1+\frac{\beta \lambda\bar{P} \nu}{ 2N_0}\right)\!.\eea}
 \end{property}
 \begin{proof}
 See Appendix A.
 \end{proof}  
Whenever $\alpha\!=\!0$, i.e., $1/\theta$ is an integer, from (\ref{Copt3}) the capacity equals 
 \bea  \label{optC} \mathcal{C}= \log\left(1+\frac{P\nu}{N_0}\right) \eea
which is the resulting capacity of a terminal if no other terminals are present with signal-to-noise ratio (SNR) equal to $P/N_0$. This is so since under such cases, $g_\ell\!=\!0$ for $\ell\!\neq\!0$. We remark that the analysis and discrete-time model of the one-dimension case is identical to that of a faster-than-Nyquist signaling system using a sinc pulse \cite{FTN1,FTN2}. 

Except for the capacity (\ref{Copt}), we are also interested in the space-normalized capacity $\bar{\mathcal{C}}$ [nats/s/Hz/m] which is defined as
\bea \label{Cbar}\bar{\mathcal{C}}=\mathcal{C}/\Delta x. \eea
With this definition, the number of signal dimensions $\rho$ in (\ref{rho}) can be calculated as the high-SNR slope of $\bar{\mathcal{C}}$,
\bea \label{rho1} \rho = \lim_{\bar{P}/N_0\to \infty} \frac{\bar{\mathcal{C}}}{\log(\bar{P}/N_0)}. \eea
Further, with the capacity in Property 1 and letting $\lambda\!\to\!0$, we have the below corollary.
\begin{corollary}
As $\lambda\!\to\!0$, for a given $\theta$ the space-normalized capacity $\bar{\mathcal{C}}$ converges to $\bar{P} \nu/N_0$ [nats/s/Hz/m].
\end{corollary}

\subsection{Capacity with Matched Filter}
Next we consider the MF capacity of each terminal corresponding to model (\ref{mod1}). That is, the capacity with only the MF applied in front, which equals
 \bea \label{Cmf} \mathcal{C}=\log\left(1+\frac{P\nu}{N_0+I}\right),  \eea
where the interference power is
 {\setlength\arraycolsep{2pt}  \bea  \label{intpow} I&=&\frac{1}{P\nu}\sum\limits_{\ell=-\infty, \ell\neq 0}^{\infty}\left|g_\ell\right|^2 \notag \\
&=&\frac{1}{\theta P\nu}\int_{-\theta/2}^{\theta/2}\!\left|G(f)\right|^2\mathrm{d}f-P\nu.\eea}
\hspace{-1.4mm}The second equality in (\ref{intpow}) is from Parseval\rq{}s identity applied to $G(f)$ in (\ref{Gf}). Following an approach similar to the one in the proof of Property 1, the interference power equals
 \bea \label{I} I=P\nu\bigg(\theta^2\left(\beta^2+2\alpha\beta+\alpha\right)-1\bigg). \eea
From (\ref{I}), under the cases that $1/\theta$ is an integer, the interference power $I\!=\!0$ and the MF capacity (\ref{Cmf}) equals the capacity of the interference-free case.

\subsection{Signal Dimensions per Meter}
With the capacities given in (\ref{Copt3}) and (\ref{Cmf}), we analyze the independent signal dimensions for the optimal and MF receivers, respectively. From (\ref{Copt3}) and (\ref{Cbar}), it can be shown that with the optimal receiver,
\bea \rho = \lim_{\bar{P}/N_0\to \infty} \frac{\bar{\mathcal{C}}}{\log(\bar{P}/N_0)} =\left\{ \begin{array}{cc}{2}/{\lambda}&\theta\geq1 \\{2\theta}/{\lambda}  &\mathrm{otherwise}. \end{array}  \right.  \notag\eea
Therefore, the maximal number of signal dimensions per meter is $2/\lambda$ for the one-dimension terminal-deployment with the optimal receiver. When $1/\theta$ is an integer, from (\ref{Cmf}) the MF can also achieve the same number of signal dimensions. 

\section{Signal Dimension Analysis for Two and Three Dimensional Cases}

\subsection{The Two-Dimensional Case: Terminals on a Plane} \label{Sec:plane}
\begin{figure}[b!]
\begin{center}
\vspace*{-6mm}
\hspace*{-2mm}
\scalebox{.35}{\includegraphics{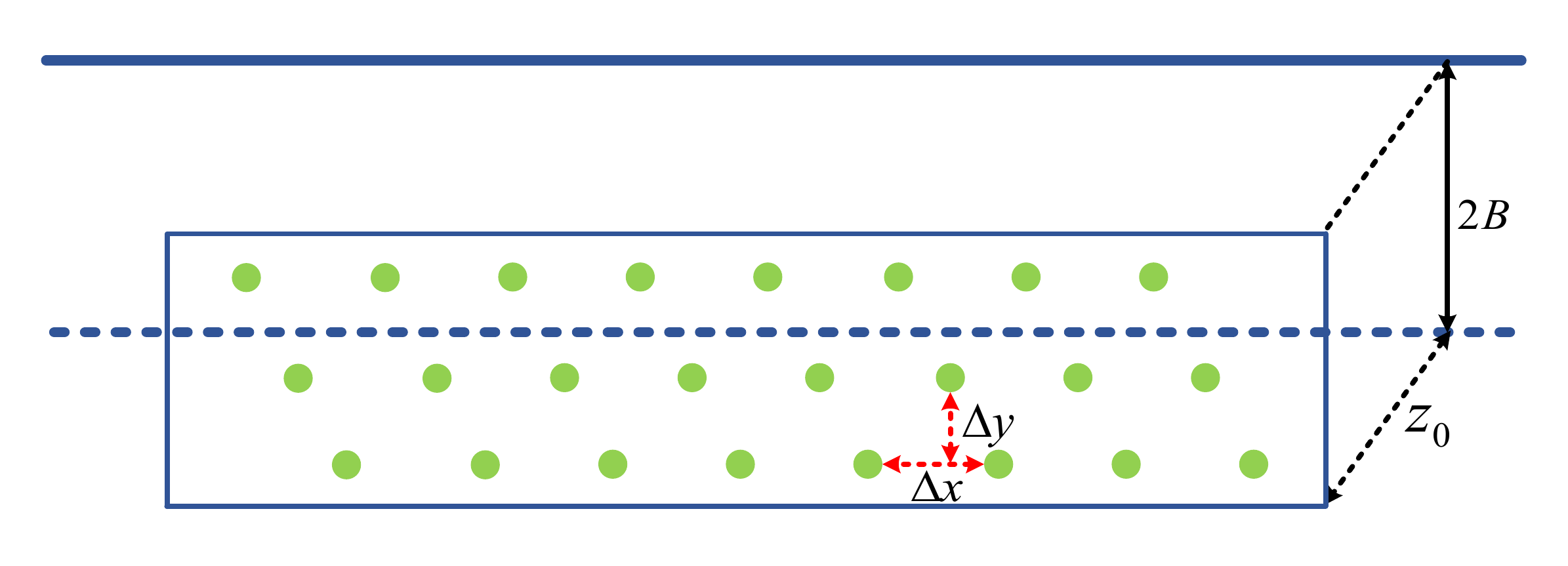}}
\vspace*{-4mm}
\caption{\label{fig3}Terminals on a plane communicating to a infinitely long wall.}
\vspace*{-2mm}
\end{center}
\end{figure}

We next move on to the case of terminals located on a two-dimensional plane at $z\!=\!z_0$, as depicted in Fig. \ref{fig3}. We are concerned with the number of independent signal dimensions per m$^2$, and we therefore let $A,B \!\to\! \infty$ to avoid edge effects. In this case, $\nu\!=\!1/2$ for all $z_0$ and capacity does not depend on distance.

The first step towards this end is to study the PSD of the signal $r(x, y)$ in the absence of noise. The PSD is given by the two-dimensional Fourier transform of the autocorrelation 
$$g(\Delta x,\Delta y) = \int_{-\infty}^\infty\int_{-\infty}^\infty s_{0,0,z_0}(x,y)s_{\Delta x,\Delta y,z_0}^\ast (x,y)\mathrm{d}x\mathrm{d}y.$$
However, under the approximation of Sec \ref{Sec:approx}, we have 
$$g(\Delta x,\Delta y) = \frac{1}{2} \mathrm{sinc}\left(\frac{2}{\lambda}\sqrt{(\Delta x)^2+(\Delta y)^2}\right).$$
As this function has radial symmetry, it follows that its Fourier transform is given by the Hankel transform\cite{H85} of degree zero, i.e.,
 {\setlength\arraycolsep{2pt}  \bea \label{Hankel} G(s) &=& 2\pi \mathcal{H}_0\{g(r)\} \nonumber \\
&=& \pi \int_{0}^\infty  \mathrm{sinc}\left(\frac{2}{\lambda}\;r \right) r{J}_0(2\pi sr)\mathrm{d}r \nonumber \\
&=& \left\{\begin{array}{ll} \frac{\lambda}{4\pi} \frac{1}{\sqrt{\frac{1}{\lambda^2}-s^2}}, & 0\leq s < \frac{1}{\lambda} \\
		 0, & s>\frac{1}{\lambda} \end{array} \right. \eea}
\hspace{-1.8mm}where $r\!=\!\sqrt{(\Delta x)^2\!+\!(\Delta y)^2}$ and ${J}_{\,0}(x)$ is the zeroth-order Bessel function of the first kind. Letting $\bar{P}$ denote the average transmitted energy per m$^2$, the normalized capacity [nats/s/Hz/m$^2$] equals
{\setlength\arraycolsep{2pt}  \bea \label{Cbar2D}  \bar{\mathcal{C}} &=&\int_{0}^{2\pi} \int_{0}^{1/\lambda} s\log\left(1+\frac{\bar{P}}{ N_0}G(s)\right) \mathrm{d}s\mathrm{d}\theta \nonumber \\
&=&\pi\left[ \frac{\log(1+\lambda N)}{\lambda^2}+N^2\log\left(\frac{N\lambda}{1+N\lambda}\right)+\frac{N}{\lambda}\right]\!, \quad\eea}
\hspace{-1.4mm}where
$$N = \frac{\lambda \bar{P}}{4\pi N_0}.$$

\begin{remark}
As $\lambda\!\to\!0$, it can be shown from (\ref{Cbar2D}) that, the limit of the space-normalized capacity $\bar{\mathcal{C}}$ equals $\bar{P}/(2N_0)$ [nats/s/Hz/m$^2$], which is the same as in the one-dimensional case with $B\!=\!\infty$.
\end{remark}
Moreover, with the normalized capacity in (\ref{Cbar2D}) we have the following property.
\begin{property}
The number of independent signal space dimensions for two-dimensional terminal-deployment equals
\bea \label{rho2} \rho = \lim_{\bar{P}/N_0\to \infty} \frac{\bar{\mathcal{C}}}{\log(\bar{P}/N_0)} = \frac{\pi}{\lambda^2}. \eea
Thus, for every $\lambda^2$ surface area deployed, we obtain $\pi$ independent signal space dimensions.
\end{property}
\begin{proof}
The result can be obtained by directly evaluating the limit in (\ref{rho2}) with $\bar{\mathcal{C}}$ in (\ref{Cbar2D}).
\end{proof}

\subsection{The Three-Dimensional Case: Terminals in Space}
From the derivations in Sec. \ref{Sec:plane}, we have already furnished for a solution of the dimensionality for the three dimensional terminal location case.
Consider the Fourier transform $S_{x_0,y_0,z_0}(\nu_1,\nu_2)$ of a signal $s_{x_0,y_0,z_0}(x,y)$. 
From the convolutional property of Hankel transforms, it follows that $G(s)$ in (\ref{Hankel}) is given by 
\bea G(s) = |S_{x_0,y_0,z_0}(s)|^2,\notag \eea
where $s\!=\!\sqrt{\nu_1^2+\nu_2^2}$. This implies that the domain of $S_{x_0,y_0,z_0}(\nu_1,\nu_2)$ is independent of the distance $z_0$ from the wall. Since the number of signal space dimensions that can be accommodated is proportional to the area of the domain of $S_{x_0,y_0,z_0}(\nu_1,\nu_2)$, it follows that the same number of dimensions is obtained in the three-dimensional case as in the two-dimensional case.

Another way to realize this result is to consider a hyper plane $\mathcal{P}=\{x,y,z\: :\: z=z_0\}$ for some small $z_0$. All signals transmitted from terminals at $z_{k}\!>\!z_0$ has to pass the plane $\mathcal{P}$. From the Huygens-Fresnel principle it, however, follows that the signal that reaches the wall can be expressed as point sources at the plane $\mathcal{P}$ that radiate the signals that reaches $\mathcal{P}$ from the terminals. That is, the signal $r(x,y)$ can be expressed as
\bea r(x,y) =\! \int_{\mathbb{R}^2} s_{\tilde{x},\tilde{y},z_0}(x,y) \sum_{k=1}^K\! \sqrt{P_k}a_k s_{x_k,y_k,z_k-z_0}(\tilde{x},\tilde{y})\mathrm{d}\tilde{x}\mathrm{d}\tilde{y}. \notag \eea
However, the number of signal space dimensions at the plane $\mathcal{P}$ is $\pi$ per $\lambda^2$ area which, means that the number of dimensions in the three-dimensional volume is unaltered compared to the two-dimensional case.

\section{Empirical Results}

\subsection{The One-Dimensional Case}

\begin{figure}[t]
\begin{center}
\vspace*{-3mm}
\hspace*{-4mm}
\scalebox{.315}{\includegraphics{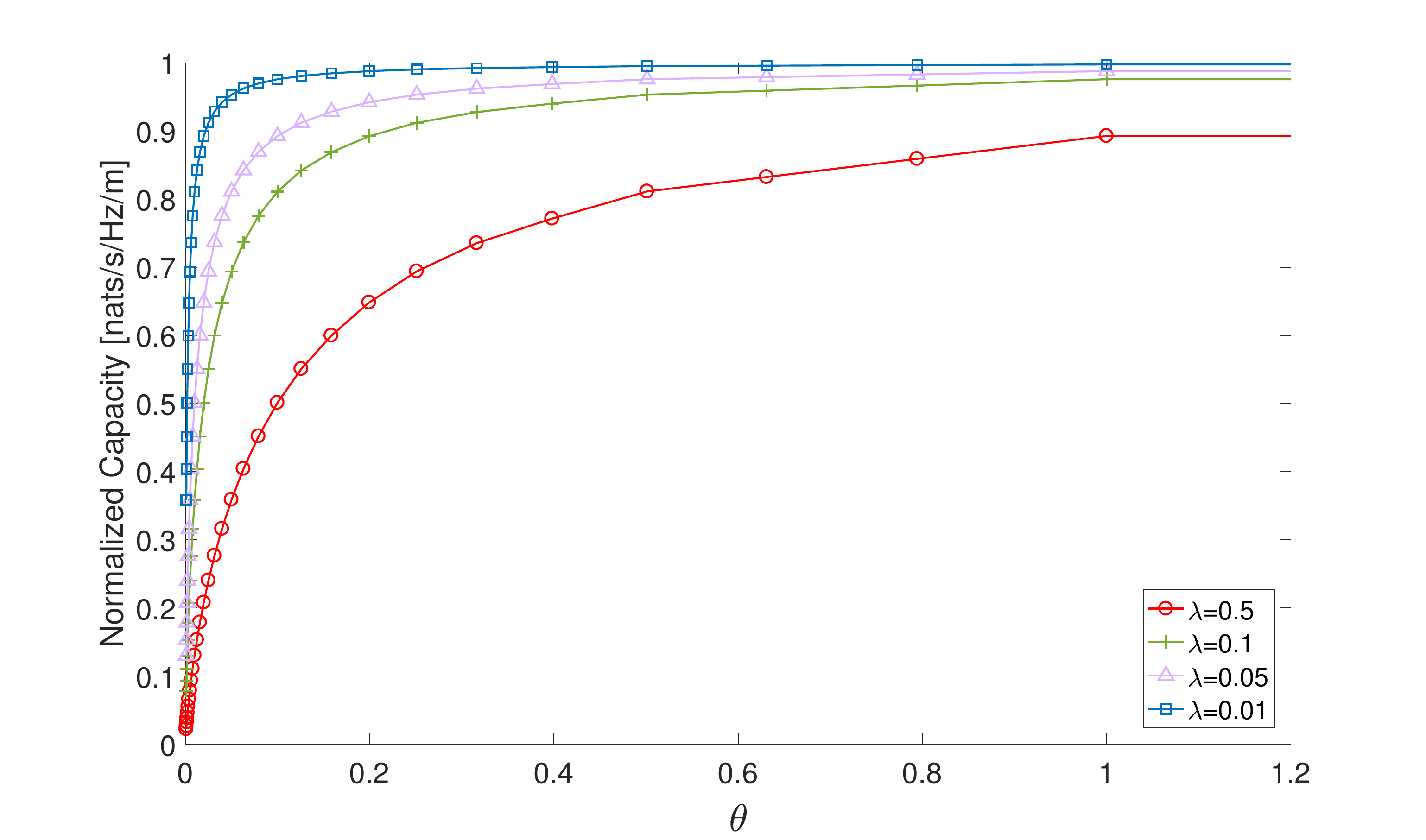}}
\vspace*{-8mm}
\caption{\label{fig4}The normalized capacity in relation to $\theta$ for optimal receiver with $N_0\!=1$, $\!\nu\!=\!0.1$, and $\bar{P}\!=\!10$.}
\vspace*{-5mm}
\end{center}
\end{figure}

In Fig. \ref{fig4}, we depict $\bar{\mathcal{C}}$ [nats/s/Hz/m] for $N_0\!=1$, $\!\nu\!=\!0.1$, and $\bar{P}\!=\!10$ and different wavelengths $\lambda$ and terminal-spacing $\Delta x$ for the optimal receiver. As can be seen, as $\lambda\!\to\! 0$, $\bar{C}$ converges to the limit $1$ in this case.

\begin{figure}
\begin{center}
\vspace*{-2mm}
\hspace*{-4mm}
\scalebox{.315}{\includegraphics{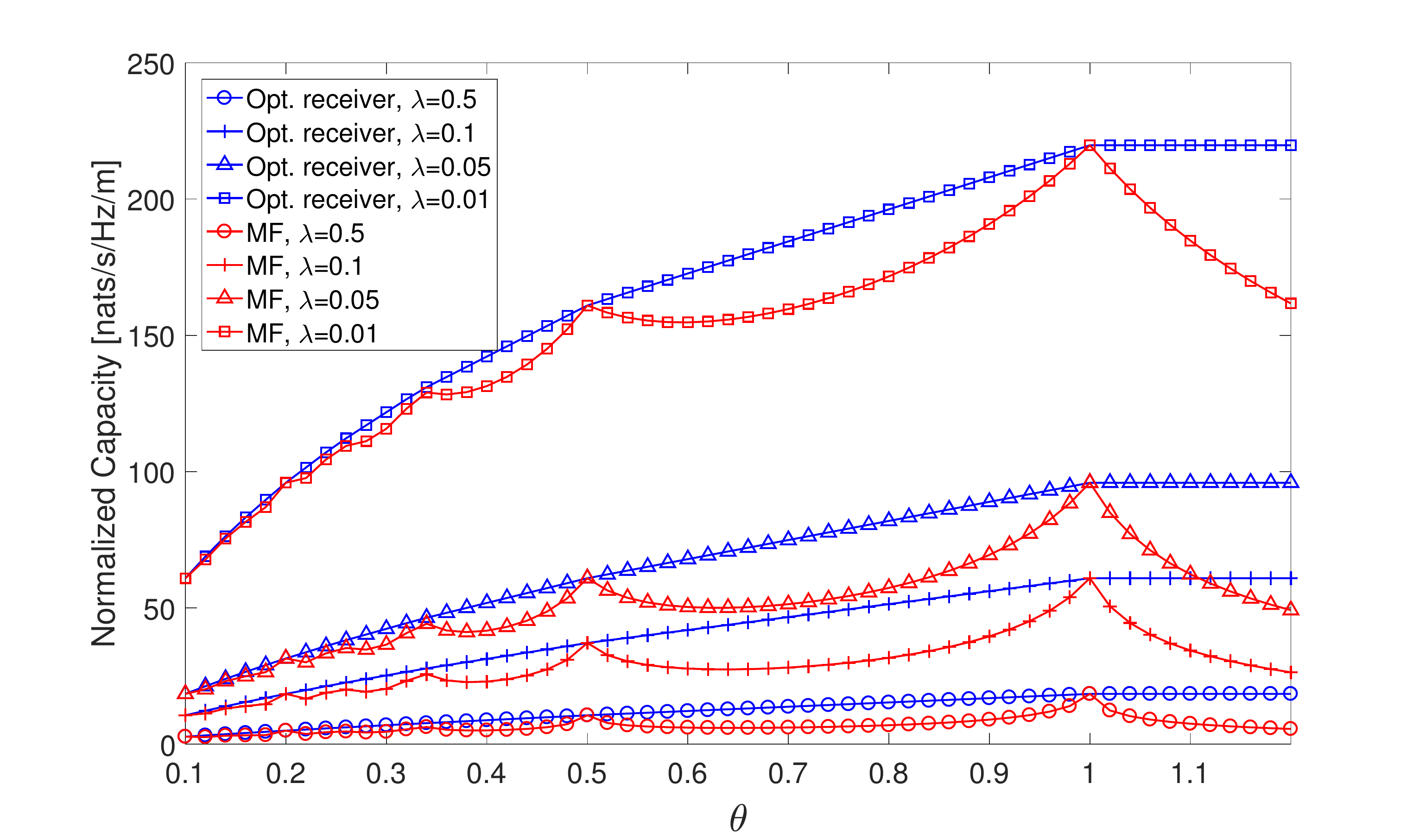}}
\vspace*{-8mm}
\caption{\label{fig6} The normalized capacity in relation to $\lambda$ and $\Delta x$ for the optimal and the MF receivers with $N_0\!=0.05$, $\!\nu\!=\!0.5$, and $\bar{P}\!=\!40$.}
\vspace*{-6mm}
\end{center}
\end{figure}

\begin{figure}[t]
\begin{center}
\vspace*{-2mm}
\hspace*{-7mm}
\scalebox{.315}{\includegraphics{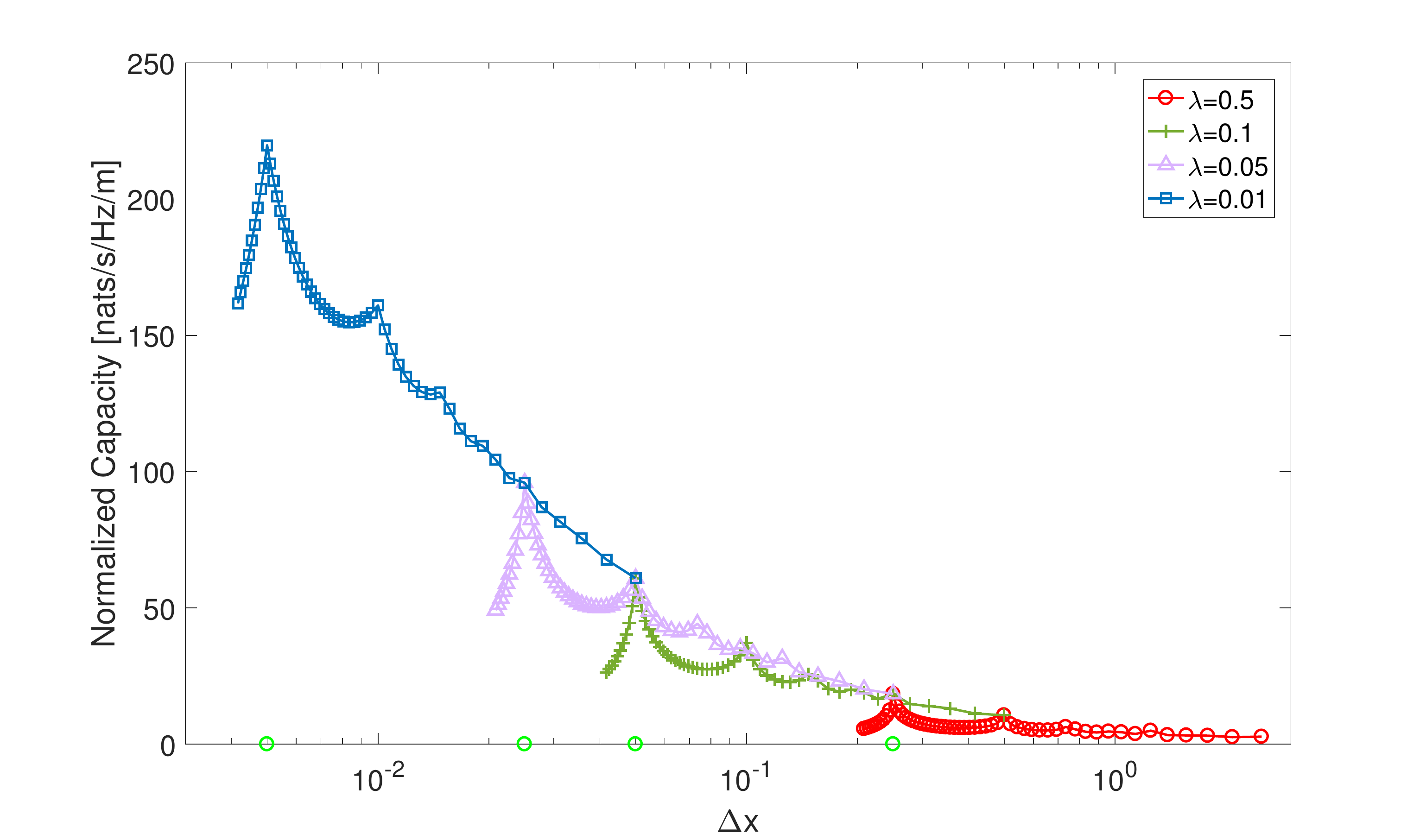}}
\vspace*{-8mm}
\caption{\label{fig7} The same test as in Fig. \ref{fig6}. The normalized capacity in relation to $\Delta x$. The green-circles correspond to $\theta\!=\!1$ for different values of $\lambda$.}
\vspace*{-4mm}
\end{center}
\end{figure}

In Fig. \ref{fig6}, we compare $\bar{\mathcal{C}}$ obtained with the optimal and the MF receivers for $N_0\!=0.05$, $\!\nu\!=\!0.5$, $\bar{P}\!=\!40$, and different $\lambda$. As can be seen, whenever $1/\theta$ is an integer, terminals do not interfere with each other and the normalized capacities of the optimal receiver and the MF are the same. In the other cases, the MF receiver is inferior to the optimal receiver. In Fig. \ref{fig7} we depict $\bar{\mathcal{C}}$, obtained with the MF as a function of terminal-distance $\Delta x$, with peaks attained when $1/\theta$ is an integer. 

\begin{figure}
\begin{center}
\vspace*{-3mm}
\hspace*{-7mm}
\scalebox{.315}{\includegraphics{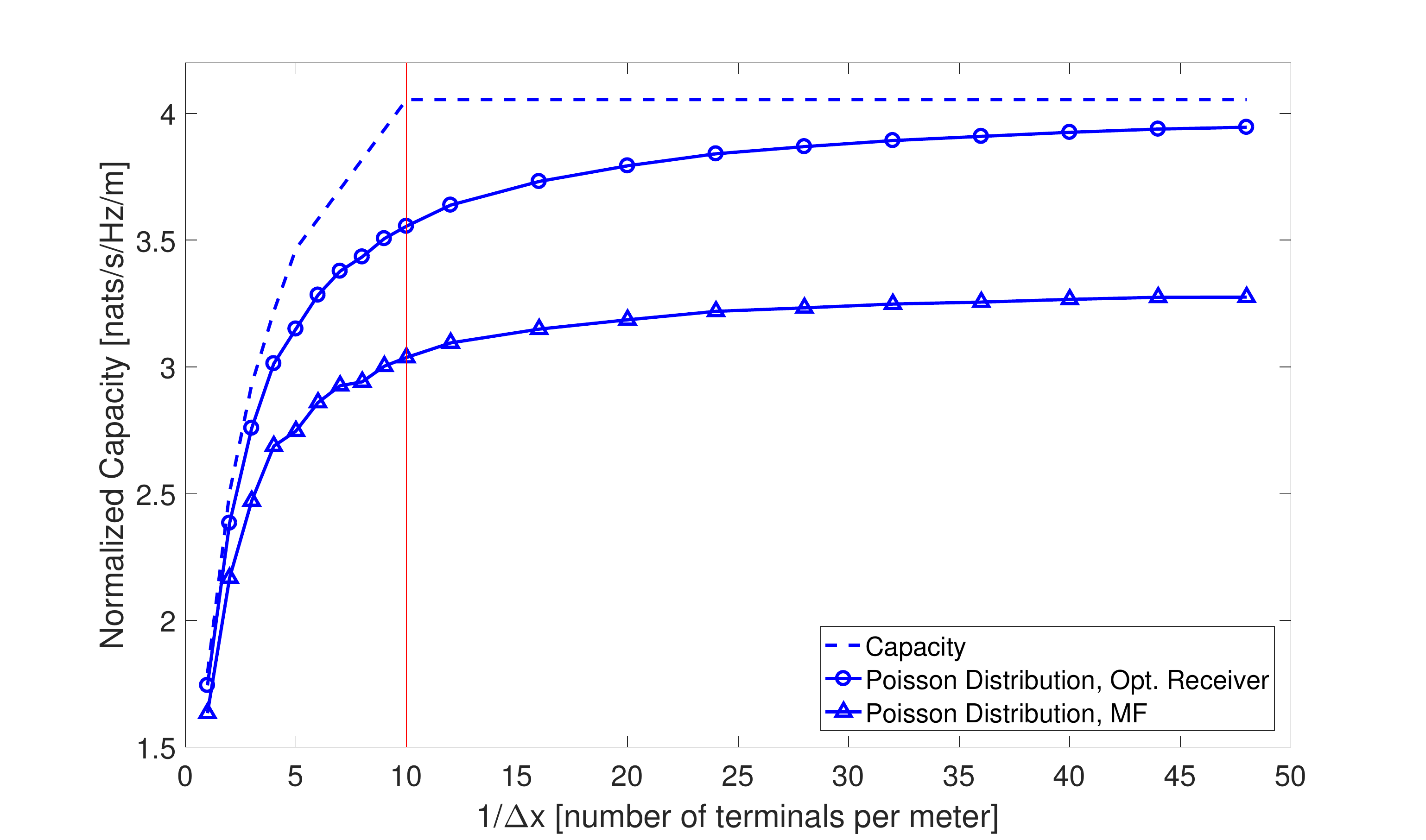}}
\vspace*{-8mm}
\caption{\label{fig8}The normalized capacity of randomly distributed terminals compared to the ideal case. We assume that $A=B=\infty$, $N_0=1$, $\bar{P}=10$, $\lambda=0.2$ and the terminals are distributed in a line with $10$m long.}
\vspace*{-7mm}
\end{center}
\end{figure}

In Fig. \ref{fig8}, we measure the normalized capacity for random allocated terminals in a 10m long line. The number of terminals are draw from a Poisson distribution for a given terminal density 1/$\Delta x$. As can be seen that, when $\Delta x$ decreases to 0, the normalized capacity reaches the capacity limit that starts to saturate at ($\Delta x\!=\!\lambda/2\!=\!0.1$) for the optimal receiver. With MF receiver, the capacity also converges when $\Delta x$ decreases and is inferior to the optimal receiver as expected.

\subsection{Two and Three Dimensional Cases}
\begin{figure}[t]
\begin{center}
\vspace*{-2mm}
\hspace*{-4mm}
\scalebox{.315}{\includegraphics{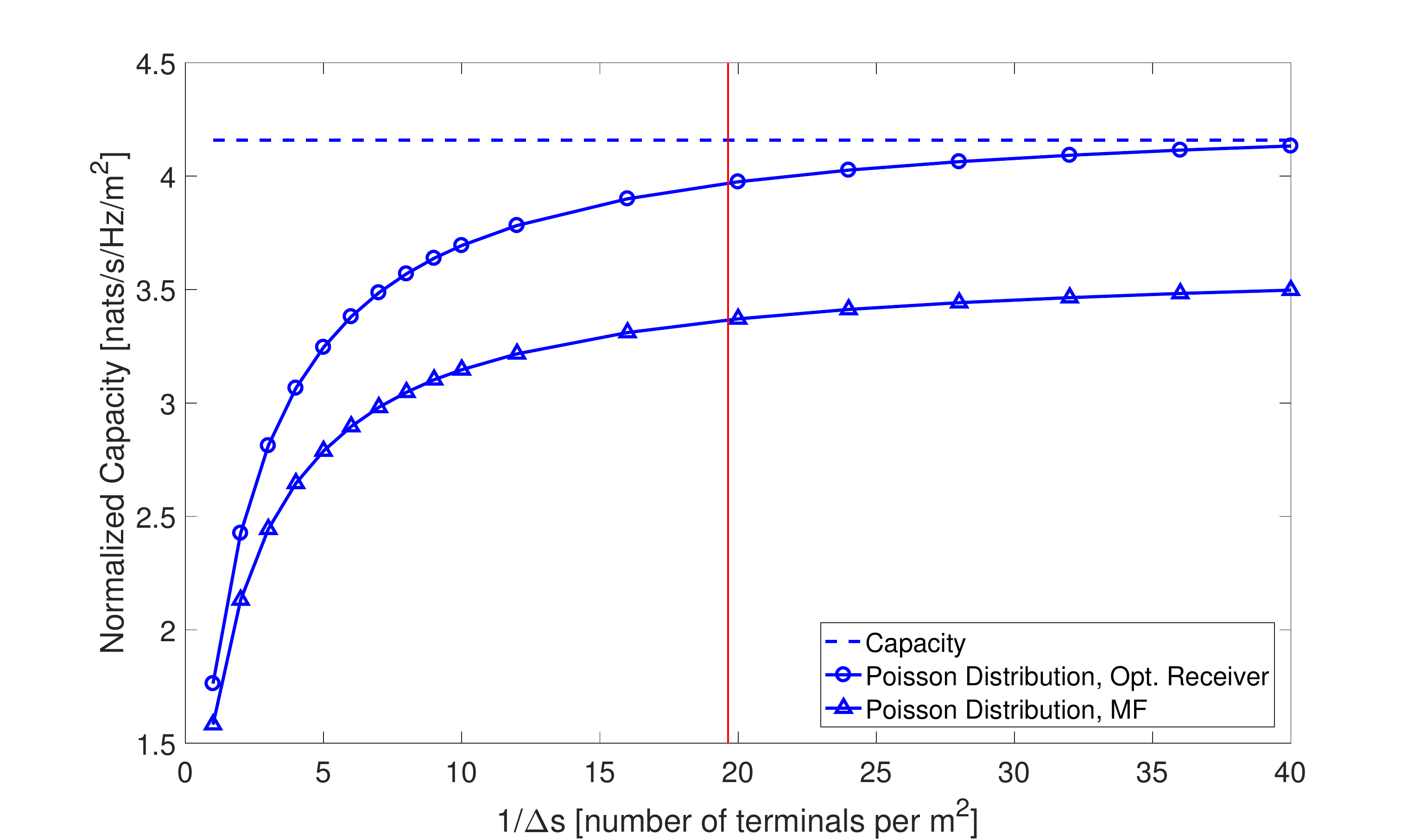}}
\vspace*{-8mm}
\caption{\label{fig9}The normalized capacity of randomly distributed terminals in a plane compared to the ideal case. We assume that $A=B=\infty$, $N_0=1$, $\bar{P}=10$, $\lambda=0.4$ and the terminals are distributed in a plane with size 20m$\times$20m.}
\vspace*{-4mm}
\end{center}
\end{figure}

In Fig. \ref{fig9}, we measure the normalized capacity for random allocated terminals in a two-dimensional plane with area 20m$\times$20m. The number of terminals are also drawn from a Poisson distribution for a given terminal density 1/$\Delta s$. As can be seen, when $\Delta s$ decreases to 0, the normalized capacity reaches the capacity limit that starts to saturate at $\Delta s\!=\!\lambda^2/\pi\!$ for the optimal receiver. With MF, the capacity also converges when $\Delta s$ decreases and is inferior to the optimal receiver.

Next we simulate the three-dimensional case, where we consider a cube with volume 4m$\times$4m$\times$4m. At the front wall of the cube, we assume that an intelligent surface with size 2m $\times$1m is deployed in the middle, for instance, we can use a white-board in a room as the surface. Since we have a surface with finite size, we use the numerical method to calculate $G_{\ell k}$ instead of using the sinc function approximation.

\begin{figure}
\begin{center}
\vspace*{-3mm}
\hspace*{-4mm}
\scalebox{.315}{\includegraphics{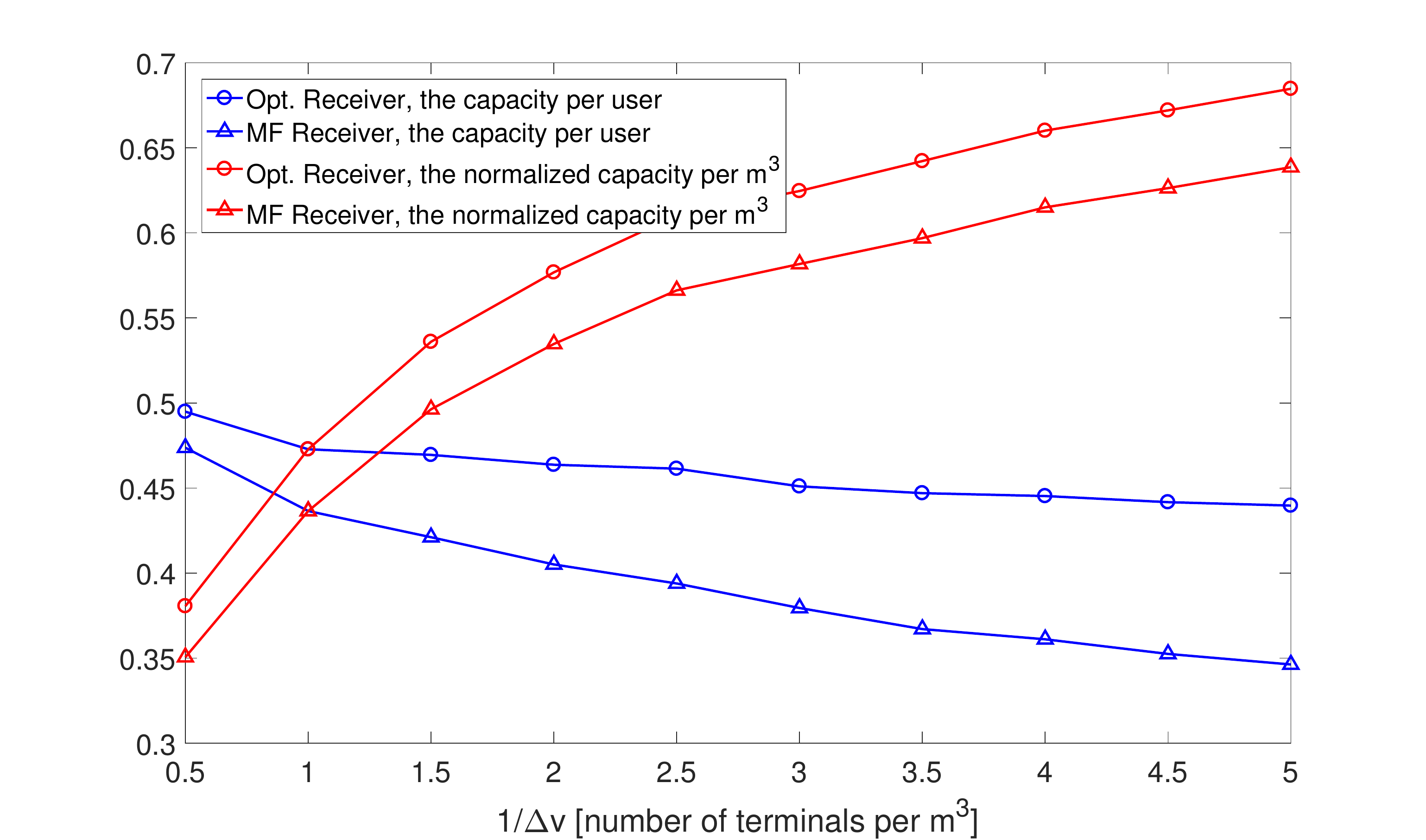}}
\vspace*{-8mm}
\caption{\label{fig11}The normalized capacity of randomly distributed terminals in a cube with volume 4m$\times$4m$\times$4m. We assume that $A\!=\!2$, $B\!=\!1$, $N_0=1$, $\bar{P}=10$ or $P=10$ and $\lambda=0.5$.}
\vspace*{-6mm}
\end{center}
\end{figure}

In Fig. \ref{fig11}, we show the space-normalized capacity for randomly located terminals in the cube. The number of terminals are also drawn from a Poisson distribution for a given terminal density 1/$\Delta v$. We consider two cases. The first case is that, we fix the transmit power of each user to be $P\!=\!10$ and then measure the capacity per user. The other case is that, we fix the power per m$^3$ to $\bar{P}\!=\!10$ and estimate the space-normalized capacity per m$^3$. As can be seen, when $\Delta v$ decreases to 0, the space-normalized capacity increases both for the optimal and MF receivers, like in the one and two dimensional cases. The capacity per user, however, is fairly flat when the number of terminals increases from 32 to 320, while the latter one results in more interferences among terminals. This clearly shows the potential of intelligent surfaces for interference suppression.

\section{Summary}
In this paper, we have considered using intelligent surfaces as large antenna array systems for communications. We have shown that, under the constraint that the transmit power per area-unit $\bar{P}$ is fixed, the normalized capacity per area-unit is $\bar{P}/(2N_0)$ when the wave-length $\lambda$ goes to zero. We have also derived that the number of independent signal dimensions per area-unit for the one-dimension case is $2/\lambda$ and $\pi/\lambda^2$ for two and three dimensional cases. In addition, we have also shown that the intelligent surfaces provide robust performance when the number of terminals increases.

\appendices
\section{Proof of Property 1}
We first define an auxiliary parameter $\tilde{\theta}\!=\!1\!-\!\beta\theta$. From the definition of $G(f)$ in (\ref{Gf}), the capacity (\ref{Copt}) can be split into two parts. In a first part, $G(f)$ is folded by $\beta$ times with amplitude $\beta\theta P\nu$ and the integration interval length being $\theta-\tilde{\theta}$, and in a second part, $G(f)$ is folded by $\beta\!+\!1$ times with amplitude $(\beta\!+\!1)\theta P\nu$ and the integration interval length being $\tilde{\theta}$. Hence, the capacity (\ref{Copt}) equals
 \be \mathcal{C}\!=\! \frac{1}{\theta}\Bigg(\!(\theta-\tilde{\theta})\log\!\left(\!1\!+\!\frac{\beta\theta P\nu }{N_0}\!\right)+\tilde{\theta}\log\!\left(\!1\!+\!\frac{(\beta\!+\!1)  \theta P\nu}{ N_0}\!\right)\!\!\Bigg)\!. \notag \ee
By the definition of $\alpha$, $\beta$ in (\ref{a}) and utilizing (\ref{theta}) yields the capacity stated in Property 1.

\bibliographystyle{IEEEtran}

\end{document}